\documentclass[a4paper,12pt]{article}
\usepackage[T1]{fontenc}
\usepackage[utf8]{inputenc}
\usepackage{lmodern}
\usepackage{fullpage}
\usepackage{amsmath}
\usepackage{amsfonts}
\usepackage{amssymb}
\usepackage{mathrsfs}
\usepackage{tensor}

\title{A gauge theory of massive spin one particles \footnote{This paper is dedicated to Prof. R. Anishetty on his sixtieth birthday, wishing him a pleasant and productive time ahead.}} 
\author{Vivek M. Vyas\\Institute of Mathematical Sciences,\\ Taramani, Chennai 600 112, INDIA \and V. Srinivasan\\Department of Theoretical Physics,\\ Guindy Campus, University of Madras, Chennai 600025, INDIA}

\begin{document}

\maketitle

\begin{abstract}%
An Abelian gauge theory describing dynamics of massive spin one bosons is constructed. This is achieved by appending to the Maxwell action, a gauge invariant mass term. The theory is quantised in temporal as well as Lorentz gauge, and the corresponding Hilbert spaces are constructed. In both the gauges, it is found that, the theory respects Lorentz invariance, locality, causality and unitarity.  
\end{abstract}

\section{Introduction}
Maxwell electrodynamics is the best known example of a theory exhibiting gauge invariance. The theory is defined by Lagrangian:
\begin{equation} \label{maxlag}
\mathscr{L} = - \frac{1}{4} F_{\mu \nu} F^{\mu \nu}, 
\end{equation}
with $F^{\mu \nu} = \partial^{\mu} A^{\nu} - \partial^{\nu} A^{\mu}$ being the field strength tensor, and $A^{\mu}$ being the Abelian gauge field. The gauge field transforms under gauge transformations as $\delta A_{\mu}(x) = \partial_{\mu} \Lambda(x)$ ($\Lambda(x)$ is some arbitrary function of coordinates), so that field strength tensor $F_{\mu \nu}$ is gauge invariant. The equations of motion that follow from above Lagrangian are given by:
\begin{equation} \label{maxeq}
\partial_{\mu} F^{\mu \nu} = 0,
\end{equation}
which are gauge invariant as well. Working in radiation gauge $\vec{\nabla} \cdot \vec{A} = 0$ with $A_{0} = 0$ it is easy to see that these equations yield a wave equation for $\vec{A}$:
\begin{equation}
\partial^{2} \vec{A} = 0.
\end{equation}
This equation with the radiation gauge condition, implies that, the Maxwell theory has two independent massless modes, which are identified with photons with two independent polarisation (spin) states \cite{greiner}. 

It is natural to ask if the Maxwell theory can be modified so as to describe massive analogues of photons having three (polarisation) spin states. This was answered affirmatively by Schwinger \cite{sch1,sch2}, who showed that gauge invariance and mass can coexist. It was showed that, if the gauge field is coupled to a conserved current $j_{\mu}$, and the theory is such that current correlator $\langle vac | j_{\mu} j_{\nu}| vac \rangle$ has a pole at $p^{2} = 0$\footnote{In many well known theories, like the BCS theory, this is achieved by spontaneous symmetry breaking \cite{nambu, ume}.}, then the gauge bosons in the theory are invariably massive \cite{sch1}. In this manner, certainly one obtains massive spin one gauge bosons but in an interacting theory. One wonders if it is possible to have a noninteracting theory of massive spin one gauge particles (that is without any coupling with any kind of current), similar in spirit as free Klein-Gordon theory or free Dirac theory.

The Proca theory is one such simple modification of Maxwell theory in this direction, and is defined by \cite{greiner, nakanishi}:
\begin{equation} \label{proca}
\mathscr{L} = - \frac{1}{4} F_{\mu \nu} F^{\mu \nu} + \frac{m^{2}}{2} A_{\mu} A^{\mu}. 
\end{equation}
The equation of motion for Proca theory is given by:
\begin{equation}
\left( \partial^{2} + m^{2} \right) A_{\mu} = 0,
\end{equation}
with the vector field obeying the condition $\partial_{\mu} A^{\mu} = 0$. These imply that the theory describes three independent massive modes, which correspond to three different spin states. However, unlike Maxwell theory, the Proca theory is not a gauge theory. Also it suffers from a problem that the Feynman propagator in this theory is not Lorentz covariant \cite{nakanishi, ruiz}. 

There exist a technique due to Stuckelberg \cite{ruiz}, to convert Proca theory into a gauge theory. The technique is to add additional field $B$ in the Proca Lagrangian, so that it now reads:
\begin{equation}
\mathscr{L} = - \frac{1}{2} F_{\mu \nu} F^{\mu \nu} + {m^{2}} \left( A_{\mu} - \frac{1}{m^{2}} \partial_{\mu} B \right)^{2} - \left( \partial_{\mu} A^{\mu} + m B \right)^{2}.  
\end{equation}
The gauge transformation are given by $\delta A_{\mu} = \partial_{\mu} \Lambda(x) $ and $\delta B = m \Lambda(x)$, where the function $\Lambda(x)$ solves $(\partial^{2} + m^{2}) \Lambda = 0$. Akin to Gupta-Blueler subsidiary condition of QED, in this theory the physical states are the ones for which $\langle phys \mid \partial_{\mu} A^{\mu} + m B \mid phys' \rangle = 0$ is obeyed.

Stuckelberg theory indeed provides one with a gauge theory that describes spin one massive modes, and is renormalisable \cite{ruiz}. However, one wonders if it possible to construct a gauge theory involving only $A_{\mu}$ field, yet describing massive unit spin modes. Takahashi and Palmer \cite{taka1} studied a gauge theory that has this desirable feature. The equation of motion of their theory is given by:
\begin{equation} \label{tp}
\left( \partial^{2} + m^{2} \right) B^{\mu \nu} = 0.
\end{equation} 
The connection with the Proca theory is made by identifying $B^{\mu \nu} = \tensor[^{\ast}]{F}{^{\mu \nu}} = \epsilon^{\mu \nu \rho \sigma} F_{\rho \sigma}$, and imposing Lorentz gauge condition $\partial_{\mu} A^{\mu}=0$. Takahashi and Palmer work with gauge invariant $B^{\mu \nu}$ field as the dynamical field, and quantise the theory by postulating commutation relation between $B^{\mu \nu}$ fields \emph{i.e.} $[B^{\mu \nu}(x), B^{\rho \sigma}(y)]$.

One way to get gauge theory with massive gauge bosons is by adding a gauge invariant mass term to Maxwell action. This possibility was  discussed by Cornwall \cite{cornwall} where the mass term of the form $F_{\mu \nu} \frac{1}{\partial^{2}} F^{\mu \nu}$ was proposed. The occurrence of operator $\frac{1}{\partial^{2}}$ in the action however lead to a presumption that it compromises locality in the theory, and as a result its proper study was not pursued. Su \cite{su} studied, both in Abelian and non-Abelian case, the possibility of adding such a gauge invariant mass to the action, in the path integral framework. However, in both these works, canonical quantisation which leads to proper identification of mode spectrum and that of the underlying Hilbert space was absent. More importantly, the essential questions regarding locality and causality remained unanswered.

In this paper, a gauge theory describing massive spin one field is constructed and quantised, with gauge field $A_{\mu}$ being the dynamical field. This is achieved by adding to the Maxwell Lagrangian (\ref{maxlag}), a gauge invariant mass term of the form $F_{\mu \nu} \frac{1}{\partial^{2}} F^{\mu \nu}$. The theory is quantised in two gauges: temporal gauge $A_{0}=0$ and Lorentz gauge $\partial_{\mu} A^{\mu} = 0$. In both the gauges, it is found that the theory respects locality, causality, Lorentz invariance and unitarity. It is found that, this theory reduces to the Proca theory in Lorentz gauge, in the classical case. In the quantum case, it is found that, one obtains a theory of massive spin one bosons, that was studied by Nakanishi \cite{nak, nakanishi}, which is free from the problem of Lorentz noninvariance, unlike Proca theory. Presence of a continuous symmetry corresponding to field redefinition, which is spontaneously broken, is found and its implications are discussed.

\section{Classical theory}

Consider a theory, dealing with an Abelian gauge field $A_{\mu}$, defined by Lagrangian:
\begin{equation} \label{Lag}
\mathscr{L} = - \frac{1}{4} F_{\mu \nu} F^{\mu \nu} - \frac{m^{2}}{4} F_{\mu \nu} \frac{1}{\partial^{2}} F^{\mu \nu}.
\end{equation}
Here, $F^{\mu \nu} = \partial^{\mu} A^{\nu} - \partial^{\nu} A^{\mu}$ is the gauge invariant field strength tensor. Appearance of differential operator $\frac{1}{\partial^{2}}$ in the Lagrangian may lead one to think that this theory violates locality and causality. However it will be seen in the next section, that none of these features are lost. Action of the operator $\frac{1}{\partial^{2}}$ can be understood by going to Fourier space {\footnote{Alternatively, one can also think of the action of this differential operator in terms of convolution by a suitable Greens function $G(x)$, which is determined by the appropriate boundary conditions of the problem. It is defined so as to solve: $\partial^{2} G(x) = \delta(x)$. Formally this implies: $G(x) = \frac{1}{\partial_{x}^{2}} \delta(x)$, so that $ \frac{1}{\partial_{x}^{2}} f(x) = \int dy  \frac{1}{\partial_{x}^{2}} \delta(x-y) f(y) = \int dy G(x-y) f(y)$. Throughout this paper, we have exploited the identity: $\partial_{x}^{2} \frac{1}{\partial_{x}^{2}} f(x) = \frac{1}{\partial_{x}^{2}} \partial_{x}^{2} f(x) = f(x)$.}}, that is:
\begin{equation}
\frac{1}{\partial^{2}} f(x) = \int \frac{d^{4}p}{(2 \pi)^{4}} \frac{-1}{p^{2}} e^{- i p x} \tilde{f}(p),
\end{equation}
where $\tilde{f}(p)$ is Fourier transform of $f(x)$ \footnote{While evaluating this integral, the input about boundary conditions needs to be given. Generally, this is given by Feynman $\epsilon$ parametrisation, for example, in some cases this is done by replacing $p^{2}$ in denominator by $p^{2} + i \epsilon$, with $\epsilon$ being infinitesimal.}. Presence of this operator in the action is not so surprising. For example, it is known to appear in the effective action of Schwinger model when one integrates out fermions \cite{das}. It also shows up in the action of two dimensional gravity theory studied by Polyakov \cite{polyakov}.  

The equations of motion that follow from above Lagrangian are given by:
\begin{equation} \label{lageqn}
\left( 1 + \frac{m^{2}}{\partial^{2}} \right) \partial_{\mu} F^{\mu \nu} = 0.
\end{equation}
This can be rewritten in terms of $\overrightarrow{E} = - \partial_{0} \overrightarrow{A} - \overrightarrow{\nabla} A_{0}$ and $\overrightarrow{B}=\overrightarrow{\nabla} \times \overrightarrow{A} $ fields as:
\begin{align} \label{gauss}
& \left( 1 + \frac{m^{2}}{\partial^{2}} \right) \: \overrightarrow{\nabla} \cdot \overrightarrow{E} = 0,\: \text{and} \\
& \left( 1 + \frac{m^{2}}{\partial^{2}} \right) \left( \overrightarrow{\nabla} \times \overrightarrow{B} - \frac{\partial \overrightarrow{E}}{\partial t} \right) = 0.
\end{align}
Two other equations obeyed by $\overrightarrow{E}$ and $\overrightarrow{B}$, which follow from Bianchi identity $\epsilon^{\mu \nu \rho \sigma} \partial_{\nu} F_{\rho \sigma}$ = 0, are:
\begin{align}
& \overrightarrow{\nabla} \times \overrightarrow{E} + \frac{\partial \overrightarrow{B}}{\partial t} = 0, \: \text{and} \\ 
& \overrightarrow{\nabla} \cdot \overrightarrow{B} = 0.
\end{align}
Combining these equations one finds that $\overrightarrow{E}$ and $\overrightarrow{B}$ fields obey a massive wave equation:
\begin{equation} \label{ebeqns}
\left( \partial^{2} + m^{2} \right) \overrightarrow{E} = 0 \quad \text{and} \quad \left( \partial^{2} + m^{2} \right) \overrightarrow{B} = 0. 
\end{equation}
These are the same as those studied by Takahashi and Palmer \cite{taka1}. The independent modes present in the theory can be easily identified by working with gauge (vector) field $A^{\mu}$, in the \emph{temporal gauge $A_{0}=0$}. In this gauge, above wave equations (\ref{ebeqns}) imply that, $\overrightarrow{A}(\vec{x},t)$ field obeys:
\begin{equation} \label{aiwaveeqn}
(\partial^{2} + m^{2} ) \overrightarrow{A}(\vec{x},t) = \overrightarrow{\nabla} f(\vec{x}), 
\end{equation} 
where $f(\vec{x})$ is some arbitrary analytic function of spatial coordinates $\vec{x}$ only. This shows that the contribution of a non-trivial $f(\vec{x})$ to $\overrightarrow{A}(\vec{x},t)$ is of non-dynamical nature. This can be easily seen by looking at the general solution of this equation:
\begin{align}
\overrightarrow{A}(\vec{x},t) = \overrightarrow{A}_{h}(\vec{x},t) + \int d^{3}y \int_{-\infty}^{\infty} dt' \: G(\vec{x},t; \vec{y},t') \overrightarrow{\nabla} f(\vec{y}),  
\end{align}    
where $G(\vec{x},t; \vec{y},t')$ is the retarded Greens function and $\overrightarrow{A}_{h}(\vec{x},t)$ is the homogeneous solution of equation (\ref{aiwaveeqn}). Owing to the fact that $(\partial_{x,t}^{2} + m^{2} ) G(\vec{x},t; \vec{y},t') = \delta (\vec{x} - \vec{x}', t - t')$, it is immediately clear that, the inhomogeneous contribution to $\overrightarrow{A}$ is independent of time, and hence would not affect the dynamics of the fields. On this ground, we set $f$ to be identically equal to zero, $f(\vec{x})=0$, henceforth. This leaves us with the homogeneous part, which obeys:
\begin{equation} \label{aiwaveeqnf}
(\partial^{2} + m^{2} ) \overrightarrow{A}(\vec{x},t) = 0,
\end{equation}
implying that the theory contains three independent massive modes, with dispersion $\omega^{2} = \vec{k}^{2} + m^{2}$. 

The same conclusion can also be inferred by working in Lorentz gauge $\partial_{\mu} A^{\mu} = 0$. From equation (\ref{lageqn}) it follows that, 
\begin{equation}
\left( \partial^{2} + m^{2} \right) A_{\mu}(x) = 0,
\end{equation}
making this theory equivalent to Proca theory, which is known to have three independent massive modes. 

In temporal gauge, the limit $m \rightarrow 0$ is well defined, and from equation (\ref{gauss}) is follows that $\overrightarrow{\nabla} \cdot \overrightarrow{A} = 0$. This eliminates one of the modes, leaving one with two massless modes of dispersion $\omega^{2} = \vec{k}^{2}$. In this manner, one obtains Maxwell electrodynamics in radiation gauge, when massless limit is considered in temporal gauge.  

\section{Quantum theory in temporal gauge}

As seen in the last section, the theory defined by (\ref{Lag}), in the temporal gauge, contains three decoupled massive modes, each obeying a Klein Gordon equation. With this observation in mind, we postulate that, the second quantised field operator $A_{i}(x)$, obeys the following commutation relation\footnote{This reminds one of the  Umezawa-Takahashi quantisation procedure, wherein unequal time commutators are used \cite{tak1, tak2, tak3}.}:
\begin{equation}
\left[ A_{i}(x), A_{j}(y) \right] = i \delta_{ij} \Delta(x - y).  \quad (i,j = 1,2,3)
\end{equation}
In above equation, both the variables $x^{\mu}$ and $y^{\mu}$ are independent in the sense that no constraints, like equal time $x_{0}=y_{0}$, exist. Here, $\Delta(x)$ is the invariant commutator function \cite{nakanishi}, defined as:
\begin{equation}
\Delta(x) = \frac{-i}{(2 \pi)^{3}} \int d^{4}p \: \text{sgn}(p_{0}) \delta(p^{2} - m^{2}) e^{-i p x}.
\end{equation}
Note that above commutation relation is compatible with the equation of motion $(\partial^{2} + m^{2} ) \overrightarrow{A}(x) = 0$. Following equal-time commutation relations, can be easily obtained from above commutation relation:
\begin{align}
& \left[ A_{i}(x), \dot{A}_{j}(y) \right]_{x_{0} = y_{0}} = i \delta_{ij} \delta(\vec{x} - \vec{y}), \: \quad \\
& \left[ E_{i}(x), B_{j}(y) \right]_{x_{0} = y_{0}} = - i \epsilon_{ijk} {\nabla}_{k}^{x} \delta(\vec{x} - \vec{y}). 
\end{align}
With this information at hand, one finds that the Hamiltonian $H$ and momentum operator $\overrightarrow{P}$ are given respectively by:
\begin{align}
& H = \int d^{3}x \: \frac{1}{2} \left( \dot{A}_{i}(x) \dot{A}_{i}(x) 
+ \overrightarrow{\nabla} A_{i}(x) \cdot \overrightarrow{\nabla} A_{i}(x) + m^{2} A_{i}(x) A_{i}(x) \right),\\  
& \overrightarrow{P} = - \int d^{3}x \: \dot{A}_{i}(x) \overrightarrow{\nabla} A_{i}(x).   
\end{align}
The quantisation of this theory in this non-covariant gauge may raise some doubts about the Lorentz covariance of the theory. It is well known that, a sufficiency condition for establishing Lorentz covariance of a given theory, is the Dirac-Schwinger covariance condition \cite{iz}, defined in terms of energy momentum tensor of the theory. The symmetric energy momentum tensor, in this case, is given by 
\begin{equation}
T^{\mu \nu} = \partial^{\mu} A_{i} \partial^{\nu} A_{i} - \frac{\eta^{\mu \nu}}{2} \left( \partial_{\rho} A_{i} \partial^{\rho} A_{i} - m^{2} A_{i} A_{i} \right).
\end{equation}
It is straightforward to check that, the Dirac-Schwinger covariance condition:
\begin{equation}
\left[ T^{00}(x), T^{00}(y) \right]_{x_{0}=y_{0}} = i \left( T^{0i}(x) + T^{0i}(y) \right) \nabla_{i}^{x} \delta(\vec{x}-\vec{y}),
\end{equation}
is indeed obeyed by $T^{\mu \nu}(x)$. 

The Fock space of this theory can be straightforwardly constructed by working with creation/annihilation operators $a^{\dagger}_{\vec{k}i}/a_{\vec{k}i}$, which are defined as:
\begin{equation}
a_{\vec{k}i} = \frac{i}{\sqrt{2 (2 \pi)^{3} \omega_{k}}} \int d^{3}x \; e^{i k x} \overleftrightarrow{\partial_{0}} A_{i}(x) \quad \text{and} \quad a^{\dagger}_{\vec{k}i} = \frac{-i}{\sqrt{2 (2 \pi)^{3} \omega_{k}}} \int d^{3}x \; e^{-i k x} \overleftrightarrow{\partial_{0}} A_{i}(x).
\end{equation} 
Note, that they obey $\left[ a_{\vec{p}i}, a^{\dagger}_{\vec{q}j} \right] = \delta_{ij} \delta(\vec{p} - \vec{q})$. This implies that, there are three massive modes, each with dispersion $\omega^{2}=\vec{k}^{2} + m^{2}$.  All of these modes are physical, in the sense that, the states describing them have positive definite norm.  

It is worth pointing out that, the general commutator of the gauge fields  $\left[ A_{i}(x), A_{j}(y) \right]$, owing to the property of invariant commutator function $\Delta(x)$, vanishes for space-like separations \emph{i.e.} when $(x-y)^{2} < 0$. This is the microcausality condition, which asserts that the theory indeed respects locality and causality \cite{iz}.  

Thus one has constructed a theory describing massive gauge particles, which are physical, and the theory respects Lorentz covariance, locality and causality.    

\section{Quantum theory in Lorentz gauge}

In many calculations it is desirable that, the theory has manifest Lorentz covariance, which is absent in above discussed temporal gauge. One of the popular gauges which possesses this property is the Lorentz gauge $\partial_{\mu} A^{\mu} = 0 $. However it is well known that, working in such a gauge leads to a quantum theory which is realised over a Hilbert space, whose metric is indefinite \cite{nakanishi}. As result, in such cases one comes across zero and negative normed states. Presence of negative normed states may result in lack of unitary time evolution in the theory, and hence their presence is not desirable. A technique to handle them, due to  Gupta and Bleuler, is to identify the physical space as a subspace of the complete Hilbert space, using a subsidiary condition. The subsidiary condition is chosen such that, the physical space does not have states which have negative definite norm \cite{greiner}. An effective way of serving this purpose is by working in the B-field formalism \cite{nakanishi}. In this formalism, one introduces a bosonic field $B(x)$ to the Lagrangian, as a Lagrange multiplier, to implement the gauge fixing constraint $\partial_{\mu} A^{\mu} = 0$. The physical subspace is subsequently identified by the subsidiary condition $B^{(+)} | phys \rangle = 0$, where $B^{(+)}$ is the positive frequency part of $B(x)$. With the addition of the $B$ field, the Lagrangian (\ref{Lag}) reads:
\begin{equation}
\mathscr{L} = - \frac{1}{4} F_{\mu \nu} F^{\mu \nu} - \frac{m^{2}}{4} F_{\mu \nu} \frac{1}{\partial^{2}} F^{\mu \nu} + B \partial_{\mu} A^{\mu}.
\end{equation}
The equations of motion that follow from above Lagrangian are:
\begin{align} 
& \left( 1 + \frac{m^{2}}{\partial^{2}} \right) \partial_{\mu} F^{\mu \nu} = \partial^{\nu} B, \\ \label{e1}
& \partial_{\mu} A^{\mu} = 0. 
\end{align}
These equations imply that, 
\begin{equation} \label{e2}
\partial^{2} B = 0, \quad \text{and} \quad \left( \partial^{2} + m^{2} \right) A^{\mu} = \partial^{\mu} B.
\end{equation}
These further imply that, 
\begin{equation} \label{e3}
\partial^{2} \left( \partial^{2} + m^{2} \right) A^{\mu} = 0. 
\end{equation}
This equation indicates that, there are both massless and massive modes present in the spectrum of the theory, described by $A_{\mu}$. Inorder to separate these two excitations, it is convenient to define two fields $V^{\mu}$ and $S^{\mu}$, as $V^{\mu} = \partial^{2} A^{\mu}$ and $S^{\mu} = A^{\mu} + \frac{1}{m^{2}} V^{\mu}$. With these definitions, one finds that,
\begin{align} \label{seqn}
\partial^{2} S^{\mu} & = 0, \quad \partial_{\mu} S^{\mu} = 0, \quad \text{and} \\ \label{veqn}
\left( \partial^{2} + m^{2} \right) V^{\mu} & = 0, \quad \partial_{\mu} V^{\mu} = 0.
\end{align} 
As done in the earlier section, the theory is quantised by postulating the commutation relation for field operator $A_{\mu}(x)$:
\begin{equation} \label{amucomm}
\left[ A_{\mu}(x), A_{\nu}(y) \right] = - i \eta_{\mu \nu} \Delta(x-y) + i \partial_{\mu}^{x} \partial_{\nu}^{x} \Delta'(x-y). 
\end{equation}
Here, $\Delta'(x)$ is defined as $\partial^{2} \Delta'(x) = \Delta(x)$, and explicitly given by \footnote{There is an ambiguity in defining $\Delta'(x)$ using $\partial^{2} \Delta'(x) = \Delta(x)$, since $\Delta'(x)$ and $\Delta'(x) + (constant) \times \Lambda(x)$ (where $\partial^{2} \Lambda(x) = 0$) both solve this equation. Requiring that commutation relation (\ref{amucomm}) is consistent with equations of motion, and with canonical equal-time commutation relation $[A_{0}(x),B(y)] = i \delta(\vec{x} - \vec{y})$, provides one with this expression for $\Delta'(x)$.}
\begin{equation}
\Delta'(x) = -\frac{1}{m^{2}} \Delta(x) - \frac{i}{m^{2}} D(x).
\end{equation}
Here, $D(x)$ is $\Delta(x)$ at zero mass \emph{i.e.} $D(x) = \Delta(x;m=0)$. 
Using equation of motion, from above commutation relation, one finds:
\begin{equation}
\left[ A_{\mu}(x), B(y) \right] = - i \left( \partial_{x}^{2} + m^{2} \right) \partial_{\mu}^{x} \Delta'(x-y). 
\end{equation} 
It must be mentioned that commutation relation (\ref{amucomm}), is compatible with equations of motion  
(\ref{e1}), (\ref{e2}) and (\ref{e3}). Using the definition of $V^{\mu}$, it is straightforward to arrive at below commutation relation from (\ref{amucomm}): 
\begin{equation}
\left[ V^{\mu}(x), V^{\nu}(y) \right] = - i m^{2} \left( \eta^{\mu \nu} m^{2} + \partial_{x}^{\mu} \partial_{x}^{\nu} \right) \Delta(x-y). 
\end{equation}
The positive/negative frequency Fourier components of $V_{\mu}$; $v_{\mu}(\vec{p})/v^{\dagger}_{\mu}(\vec{p})$, are defined as:
\begin{align}
v^{\mu}(\vec{p}) &= \frac{i}{\sqrt{2 (2 \pi)^{3} \omega_{k}}} \int d^{3}x \; e^{i k x} \overleftrightarrow{\partial_{0}} V^{\mu}(x), \: \text{and} \\
v^{\dagger \mu}(\vec{p}) &= \frac{-i}{\sqrt{2 (2 \pi)^{3} \omega_{k}}} \int d^{3}x \; e^{-i k x} \overleftrightarrow{\partial_{0}} V^{\mu}(x). 
\end{align}
with $\omega^{2}_{k}=\vec{k}^{2} + m^{2}$. Commutation relation between them are straightforwardly obtained as:
\begin{equation}
\left[ v_{\mu}(\vec{p}), v^{\dagger}_{\nu}(\vec{q}) \right] = m^{2} \left( p_{\mu} p_{\nu} - m^{2} \eta_{\mu \nu} \right) \delta(\vec{p} - \vec{q}).
\end{equation}
Without loss of generality, one can consider that one is working in a coordinate system where $p^{\mu} = (E=\sqrt{p^{2} + m^{2}},0,0,p)$. Then owing to the constraint $p_{\mu} v^{\mu}(\vec{p}) = 0$, one finds that $v^{0}(\vec{p})$ can rewritten in terms of $v^{3}(\vec{p})$: $v^{0}(\vec{p}) = \frac{p}{E} v^{3}(\vec{p})$. This implies that there are three independent massive modes, in the theory:
\begin{align}
& \left[ v_{1,2}(\vec{p}), v^{\dagger}_{1,2}(\vec{q}) \right] = m^{4} \delta(\vec{p} - \vec{q}), \\
& \left[ v_{3}(\vec{p}), v^{\dagger}_{3}(\vec{q}) \right] = m^{2} E^{2} \delta(\vec{p} - \vec{q}).
\end{align}
From above commutation relations it is clear that the states created by these operators will be positive normed states, implying that all the three massive modes are physical. It is worth pointing out that, the commutation relations and equations of motion obeyed by $V^{\mu}(x)$, are identical to a vector field theory describing massive spin one particles, which was studied by Nakanishi \cite{nak}.   

Similarly, from the commutation relation (\ref{amucomm}), one finds that commutation relation for $S^{\mu}$  is:
\begin{equation}
\left[ S^{\mu}(x), S^{\nu}(y) \right] = \frac{i}{m^{2}} \partial_{x}^{\mu} \partial_{x}^{\nu} \left( \partial^{2} + m^{2} \right) \Delta'(x-y). 
\end{equation}
Following the above treatment, one can write commutation relation for positive and negative frequency Fourier components of $S^{\mu}$:
\begin{equation}
\left[ s_{\mu}(\vec{p}), s^{\dagger}_{\nu}(\vec{q}) \right] = - \frac{p_{\mu} p_{\nu}}{m^{2}} \delta(\vec{p} - \vec{q}).
\end{equation} 
Note that in this case, since $S_{\mu}$ field is massless, so the dispersion is $p^{2} = 0$. In the case, when $p^{\mu}=(p_{3},0,0,p_{3})$, one immediately finds:
\begin{align}
& \left[ s_{1,2}(\vec{p}), s^{\dagger}_{1,2}(\vec{q}) \right] = 0, \\
& s_{3}(\vec{p}) = s_{0}(\vec{p}), \\
& \left[ s_{3}(\vec{p}), s^{\dagger}_{3}(\vec{q}) \right] = - \frac{p_{3}^{2}}{m^{2}} \delta(\vec{p} - \vec{q}). 
\end{align}
From these commutation relations, it is clear that, the states created by $s^{\dagger}_{1,2}$ are zero normed states. Whereas the ones created by $s^{\dagger}_{0,3}$ are negative normed states. Presence of these negative normed states, may raise questions about the unitarity of the theory. However, it is worth recollecting that, in this formalism, the physical states are the ones which are annihilated by the positive frequency part of B: $B^{(+)} |phys \rangle = 0$ or equivalently $b(\vec{p}) |phys \rangle = 0$. From equation (\ref{e2}), one has $S^{\mu} = \frac{1}{m^{2}} \partial^{\mu} B$, which implies that for physical states: $s_{0}(\vec{p}) |phys \rangle = 0$.  
This clearly shows that, the physical subspace of the theory does not have any negative normed states.

In this manner, one finds that, the quantum theory defined by (\ref{Lag}), in the Lorentz gauge, possesses three massive (physical) modes, which is in agreement with the results of temporal gauge obtained in the earlier section. Note that the quantum theory obtained is local, causal, manifestly Lorentz invariant and respects unitarity.    

\section{Symmetries and its consequences}

Consider the case of massless scalar theory, governed by the equation of motion:
$\partial^{2} \phi = 0$. This equation of motion is invariant under continuous field transformation $\phi \rightarrow \phi + \text{constant}$, which is a symmetry of the theory. Owing to canonical (equal-time) commutation relation: $\left[ \phi(x), \dot{\phi}(y) \right] = i \delta(\vec{x} - \vec{y})$, it is straightforward to see that charge $Q = - i \int dV \: \dot{\phi}$ generates this symmetry. 

It is well known that, in quantum field theory, symmetries can be realised in two modes: Wigner mode and Nambu mode. In Wigner mode, the generator of symmetry annihilates the vacuum: $Q |vac \rangle = 0$. On the other hand, in Nambu mode vacuum is not annihilated by symmetry generator: $Q |vac \rangle \neq 0$, and one says that the corresponding symmetry has been spontaneously broken \cite{nakanishi}. The celebrated Goldstone theorem states that, when a system realises a continuous symmetry in Nambu mode, there exists gapless (massless) Nambu-Goldstone modes in the theory. An economical way of checking whether a symmetry is realised in Wigner or Nambu mode, 
further in case, if it is realised in Nambu mode, then identifying the Nambu-Goldstone mode; is by looking at vacuum expectation of commutator of charge with any dynamical field $\Phi$ in the theory 
$\langle vac | \left[ Q, \Phi \right] | vac \rangle$ \cite{nakanishi, ume}. If this vacuum expectation is vanishing then the symmetry is realised in Wigner mode, in the given vacuum. On the otherhand, if for any dynamical field, the expectation value is non-vanishing, then it implies that symmetry generated by charge $Q$ is realised in Nambu mode (in the given vacuum), and the dynamical field itself represents the gapless Nambu-Goldstone mode \cite{ume}.  

In the above scalar field example, one finds that, the commutator itself is non-vanishing $\left[ \phi, Q \right] \neq 0$. This implies that, in all possible vacua, the symmetry generated by $Q$ is realised in Nambu mode. One readily sees that, field $\phi$ itself represents the gapless Nambu-Goldstone mode. This clearly shows that the existence of field shifting symmetry $\delta \phi = \text{constant}$, leads to massless nature of scalar particles of field $\phi$.

In the case of Maxwell theory, there exist gauge invariance in the theory: $\delta A_{\mu}(x) = \partial_{\mu} \lambda(x)$, $\lambda(x)$ being some arbitrary function of coordinates. The theory certainly possesses the field shifting symmetry $\delta A_{\mu} = \text{constant}$, and hence there must be Nambu-Goldstone modes in the theory. This aspect of the theory, can be most conveniently studied by working in Lorentz gauge, because it respects locality, an essential assumption of the Goldstone theorem. It is easy to see that, even after imposing the Lorentz gauge condition: $\partial_{\mu} A^{\mu} = 0$, the theory is still invariant under field shifting $\delta A_{\mu} = \text{constant}$. So, as per the Goldstone theorem, it follows that there exists four Nambu-Goldstone modes in the theory, since each field component can be shifted independently. However, Maxwell theory quantised in Lorentz gauge invariably has indefinite metric, and it turns out that only two transverse components of $A_{\mu}$ create massless physical particle states with positive norm (which correspond to photons with two polarisation states). In this manner, one sees that photons are actually Nambu-Goldstone modes \cite{fer, sen, brandt}.           

As in case of electrodynamics, the theory defined by (\ref{Lag}) that describes massive spin one gauge bosons, has gauge invariance, as also the field shifting symmetry $\delta A_{\mu} = \text{constant}$. In this light, one wonders about the existence of Nambu-Goldstone modes. Note that the field shifting symmetry is retained even after quantising the theory in Lorentz gauge, as seen in (\ref{e3}). From equations (\ref{seqn}) and (\ref{veqn}), it is immediately clear that, the field symmetry $\delta A_{\mu} = \text{constant}$, manifests as field shifting symmetry of $S_{\mu}$, $\delta S_{\mu} = \text{constant}$. The equal time commutation relation: 
\begin{equation}
\left[ S_{0}(x), B(y) \right]_{x_{0} = y_{0}} = i \delta(\vec{x} - \vec{y}), 
\end{equation}
allows one to define a conserved charge $Q_{B} = - i \int d^{3}x \: B(x)$, which generates the  symmetry: $\delta S_{0}(x) = - i \theta \left[Q_{B}, S_{0}(x) \right]$. From this equation, it is clear that this symmetry is realised in Nambu mode in all possible vacua. As a result of this, $S_{0}(x)$ possesses massless Nambu-Goldstone modes, created by $s^{\dagger}_{0}$. However, as noted above, these Nambu-Goldstone particle states are the ones that have negative norm, and from subsidiary condition it follows that they are unphysical. The same also holds for those created by $s^{\dagger}_{3}$. The particle states created by $s^{\dagger}_{1,2}$ are also unobservable since they have zero norm. 

\section{Conclusion}

In this paper, an Abelian gauge theory describing massive particles of unit spin is studied. It is quantised in two gauges, temporal gauge and Lorentz gauge. In both the gauges, it is found that, the theory possesses three massive modes; and respects locality, causality, Lorentz invariance and unitarity. The quantisation procedure, using unequal time commutation relations, is employed to quantise the theory in both the gauges. In Lorentz gauge, B-field formalism is used to facilitate identification of physical modes of the theory. Presence of a continuous field redefinition symmetry in Lorentz gauge, and its spontaneous breaking, is found and its implications are discussed.

\section*{Acknowledgment}
VMV thanks Prof. P. K. Panigrahi for insightful discussions regarding this work. 

\appendix
\section*{Appendix: Non-Abelian generalisation}
It is a natural question to ask whether above mentioned procedure for describing massive Abelian gauge particles can be extended to non-Abelian case. It turns out that it is possible to construct an equation which is a non-Abelian generalisation of (\ref{tp}). As mentioned above, Takahashi and Palmer propose:
\begin{equation}
  \left( \partial^2 + m^2 \right) \epsilon^{\mu \nu \rho \sigma} F_{\mu \nu} = 0,
\end{equation}
as the equation for an Abelian gauge theory describing massive spin one bosons. This equation can be generalised for non-Abelian case, as:
\begin{equation} \label{eom}
  \left( D_{\mu} \cdot D^{\mu} + m^2 \right) \epsilon^{\mu \nu \rho \sigma} F_{\mu \nu}= 0,
\end{equation}
where $F_{\mu \nu} = \partial_{\mu} A_{\nu} - \partial_{\nu} A_{\mu} - [A_{\mu}, A_{\nu}]$ is the field strength tensor for the non-Abelian gauge field $A_{\mu}$. Here, $A_{\mu}(x) = A_{\mu}^{a}(x) t^a$, where matrices obey commutation relation:
\begin{equation}
  \left[ t^a, t^b \right] = f^{abc} t^{c}.
\end{equation}   
Under a gauge transformation $U(x) = e^{i \theta_a(x) t_a}$, the gauge field $A_{\mu}$ and covariant derivative $D_{\mu} = \partial_{\mu} - i A_{\mu}$, transform as:
\begin{equation}
  {A'}_{\mu} = U A_{\mu} U^{-1} + (\partial_{\mu} U) U^{-1}, \quad {D'}_{\mu} = U D_{\mu} U^{-1}.
\end{equation}   
Field strength transforms as: ${F'}_{\mu \nu} = U F_{\mu \nu} U^{-1}$, and obeys the Bianchi identity:
\begin{equation}
  D_{\mu} F_{\nu \rho} + D_{\nu} F_{\rho \mu} + D_{\rho} F_{\mu \nu} = 0.
\end{equation}   
Unlike Abelian case discussed earlier, in this case since the equation of motion (\ref{eom}) is a nonlinear equation, above discussed procedure, involving unequal time commutators, can not be employed for quantisation. Until and unless, one constructs the Hilbert space for the theory and shows that it is a unitary representation of Poincare group, it is not possible to come to any conclusion about mass and spin of the particles in the theory.

\bibliographystyle{plain}
\bibliography{ref}

\end{document}